\documentstyle[11pt,newpasp,twoside,epsfig]{article}
\reversemarginpar

\begin{document}
\title{The Interesting Dynamics of the 55 Cancri System}
\author{Gregory S. Novak, David Lai, \& D. N. C. Lin}
\affil{UCO/Lick Observatory, University of California at Santa Cruz,
  Santa Cruz, CA 95064}

\begin{abstract}
  Recent studies of the 55 Cancri system suggest the existence of
  three planets with periods of $\sim$15 days, $\sim$45 days, and
  $\sim$5500 days (Marcy et al. 2002).  The inner two planets are near
  the 3:1 mean motion commensurability and it is likely that these two
  planets became trapped in the resonance while farther from the star
  and migrated together.  As the innermost planet begins to dissipate
  energy through tides the planets break out of the resonance.  The
  final state of the system gives important information about its past
  history, such as the migration timescale that led to capture.
\end{abstract}

\section{Introduction}

Tidal evolution leading to capture into resonance has been invoked to
explain the many commensurabilites in our solar system (e.g. Goldreich
1965), but the problem of breaking resonance due to tidal effects is
less well-studied.  The presence of even a small amount of dissipation
can alter the dynamical evolution of the system.  In this case, the
dissipation causes the resonance to accelerate the circularization of
the inner planet compared to what would happen in isolation.

The resonant interaction between the planets in the 55 Cancri system
imply that fully self-consistent fits are necessary to obtain the
orbital elements of the system.  The 3:1 resonance covers a narrow
range of semi-major axes, so determining the location of the system
within the resonance will require accurate fits, which requires
additional data.  Direct integrations of the orbital elements given by
one fully self-consistent fit (Laughlin 2002) indicate that the
resonant angles are circulating and the system in {\em not} in
resonance.

To study the long term dynamics of the system, we use the classical
disturbing function together with Lagrange's planetary equations to
lowest order in eccentricity (e.g. Murray \& Dermott 1999).  We
numerically integrate these equations using a variable time step
Bulirsch-Stoer integrator.

We consider tidal evolution in the regime where energy is dissipated
and angular momentum is conserved.  The eccentricity damping rate is
given by $\dot{e}=-e/\tau_e$ and $\tau_e=GMme^2Q/anE_0$ where $G$ is
the gravitational constant, $M$ is the mass of the star, $m$ is the
mass of the inner planet, $Q$ parameterized tidal energy dissipation,
$a$ is the planet's semi-major axis, $n$ is the planet's mean motion,
and $E_0$ is the maximum amount of energy stored in the tidal
deformation of the planet.  The planet's eccentricity e-folding time
in isolation is given by $\tau_e$.

\section{Restricted three body problem}

In order to develop an understanding of the problem, we study a
simplified case: the circular restricted three body problem where the
inner planet is massless and the outer planet is $1\ M_J$.  The
resonant libration timescale is $\sim100$ orbits of the outer planet,
the secular evolution time is $\sim3500$ orbits.  We assume that tidal
dissipation occurs only in the inner planet.  The tidal evolution
timescale is typically {\em much} longer than the dynamical and
secular timescales, so we choose a small value for Q.  This does not
change the character of the dynamics as long as the tidal timescale
remains long compared to other timescales.

\begin{figure}[htbp]
  \plottwo{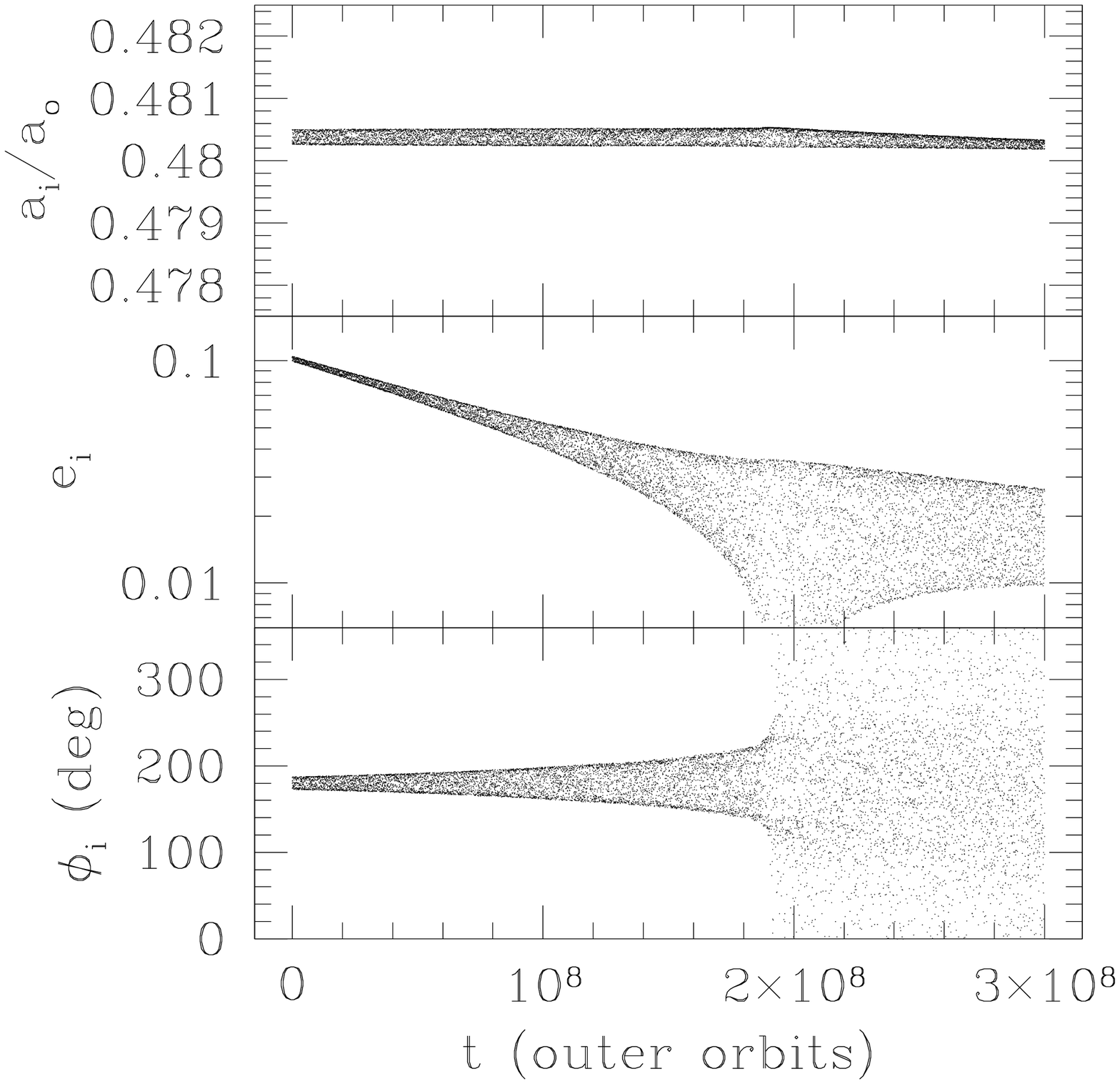}{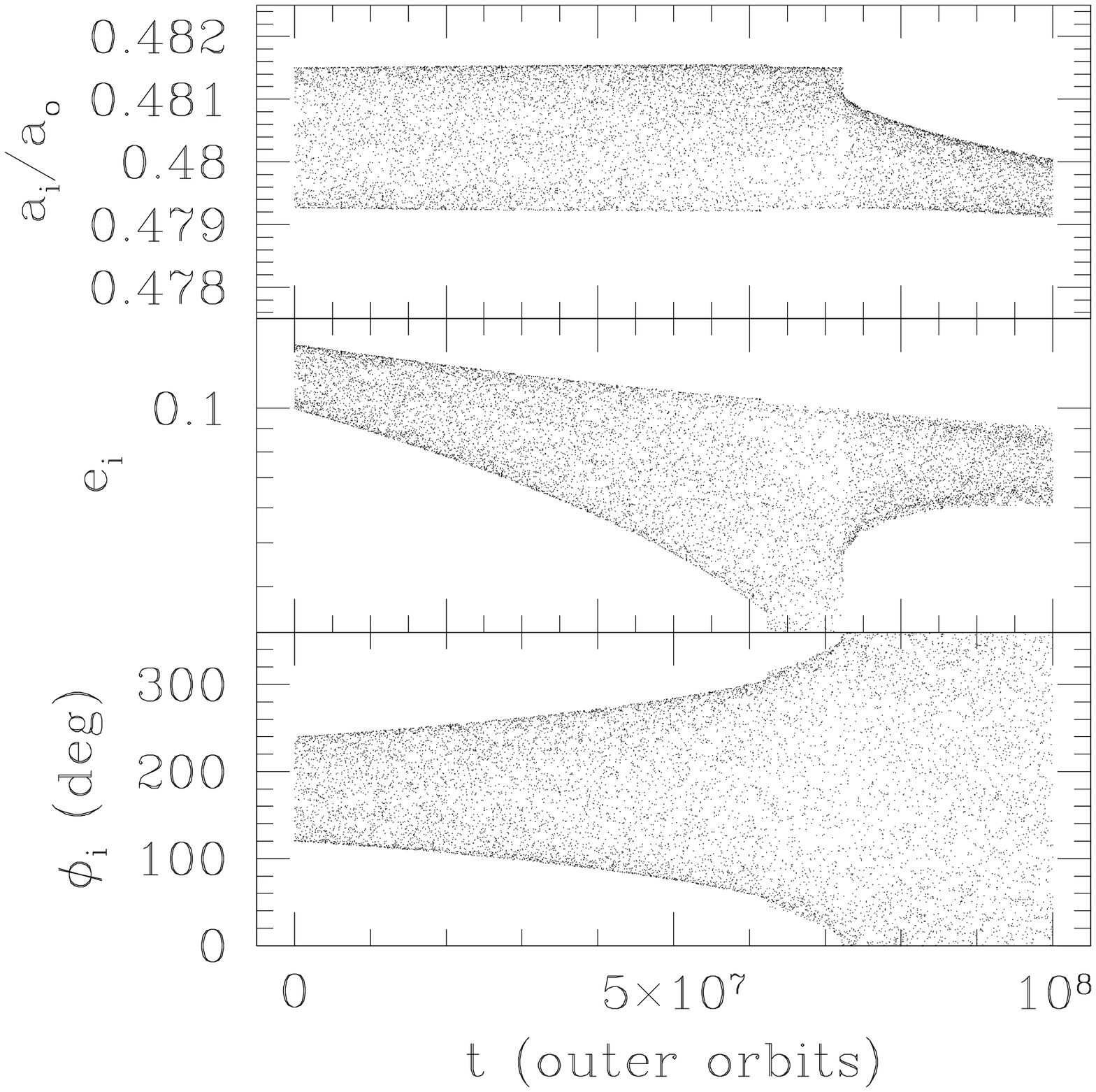}
  \caption{The time evolution of the inner planet under the influence
    of tidal dissipation in the circular restricted three body
    problem.  On the left, the system starts deep in the 3:1
    resonance, and on the right, the system is initially far from the
    exact resonance.  The outer planet supplies energy and angular
    momentum to the inner planet, resulting in an {\em accelerated}
    circularization of the inner planet compared to what would happen
    if the inner planet were isolated.}
  \label{time}
\end{figure}

As tides remove energy the outer planet supplies the inner planet with
enough energy to maintain the commensurability, but an adiabatic
invariant demands that the resonant angle's libration amplitude increase
as the eccentricity decreases, eventually breaking the resonance.  The
energy flow is arranged by forcing the resonant angle librate to about
a mean value slightly displaced from its equilibrium value of $\pi$.
The flow of angular momentum between the planets is determined by the
same resonant angle and this results in {\em faster} circularization
of the inner planet's orbit compared to the situation where the inner
planet is isolated.  In the case of the 3:1 resonance, the inner
planet circularizes five times faster in the presence of an outer
companion.

\section{Breaking the resonance}

The resonant angle's libration amplitude can be written as 

\begin{equation}
  \Delta a_i = 4 a_i \sqrt{C m_o a_i/3 M a_o} e_i^{j/2} \sin(\Delta \phi/4)
  \label{libration}
\end{equation}
where the subscripts $i,o$ refer to the inner and outer planets, $m$
is the mass of a planet, $M$ is the mass of the star, $j$ is the order
of the resonance, $\phi$ is the resonant angle, $C$ is a constant
determined by the resonance under consideration, and $\Delta a$,
$\Delta\phi$ refer to the full width of the libration. As energy is
removed, the only one of these factors to change is the inner planets
eccentricity which determines that the resonant angle will begin to
circulate when:
\begin{equation}
  e_{break} = e_{init} (\sin(\Delta\phi_{init}/4))^{2/j}
  \label{ebreakeqn}
\end{equation}
where the subscript $init$ refers to the values when tidal dissipation
became important.  Once the resonance is broken the inner planet
evolves to a circular orbit only under the influence of tides, where
the final semi-major axis is $a_{final} = a_{res}(1-e_{break}^2)$ and
$a_{res}$ is that corresponding to the exact commensurability.  The
final period ratio of the two planets gives a constraint on how deep
the two planets were in resonance when tidal effects became important.

Knowing the eccentricity decay timescale as augmented by the dynamics
of the resonance allows us to write the time required to break the
resonance after tidal effects become important as:
\begin{equation}
  \tau_{break} = -2(1+2j) \ln(\sin(\Delta\phi/4))/j
  \label{tbreakeqn}
\end{equation}

\begin{figure}[htbp]
  \plottwo{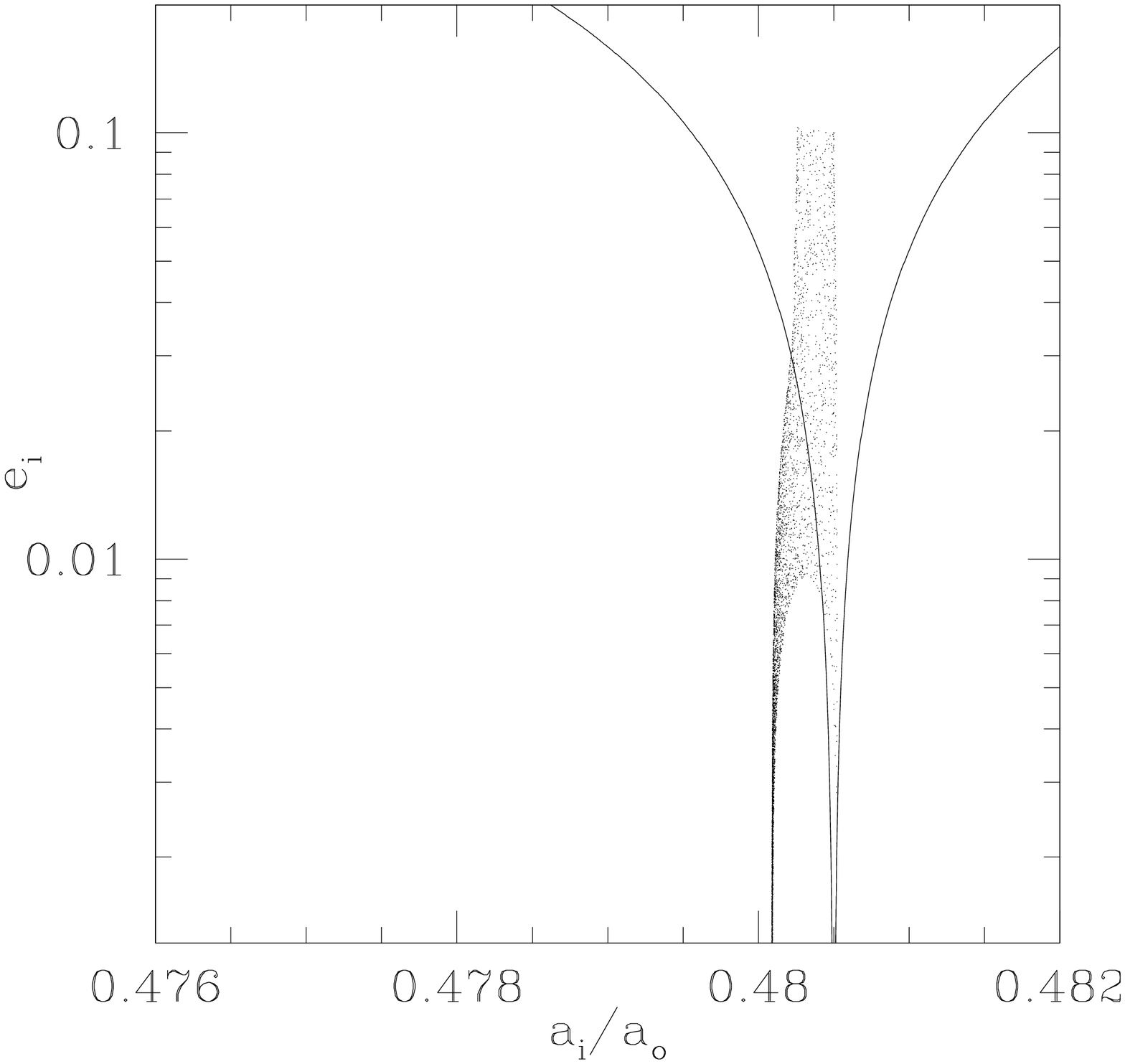}{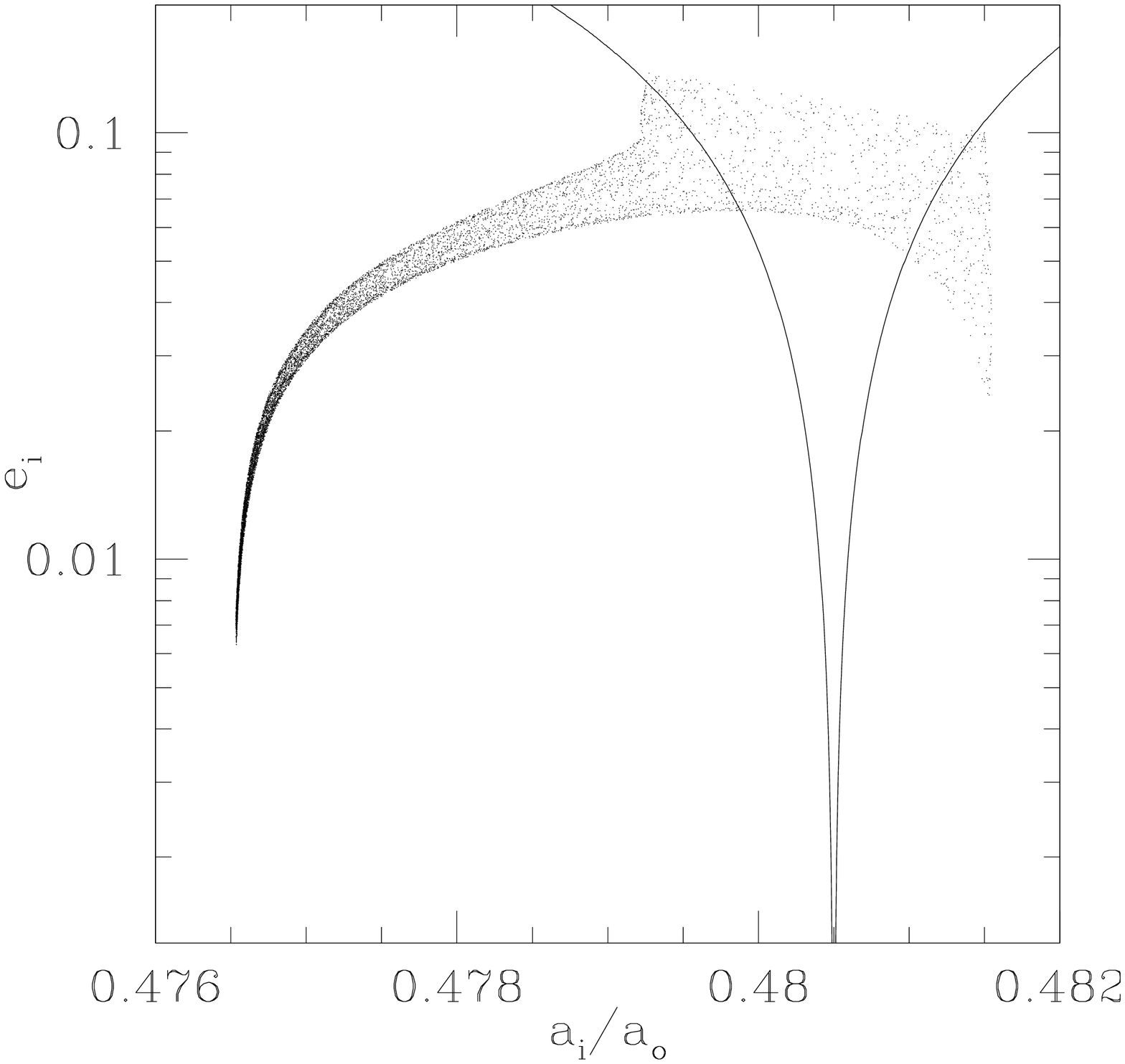}
  \caption{Semi-major axis and eccentricity as the system evolves out
    of resonance.  The lines indicate the maximum semi-major axis
    libration while in resonance.  In both plots, the planet's
    eccentricity is initially high and it evolves toward zero.  The
    plot on the left is a system that starts near the exact
    commensurability.  The inner planet has a very low eccentricity
    when the resonance is finally broken.  The system plotted on the
    left starts far from the exact resonance and the breaking occurs
    while the inner planet still has substantial $e_i$.  The
    asymptotic period ratio ($P_i/P_o$) is substantially smaller than
    the previous case..}
\label {ae}            
\end{figure}

\begin{figure}
  \plottwo{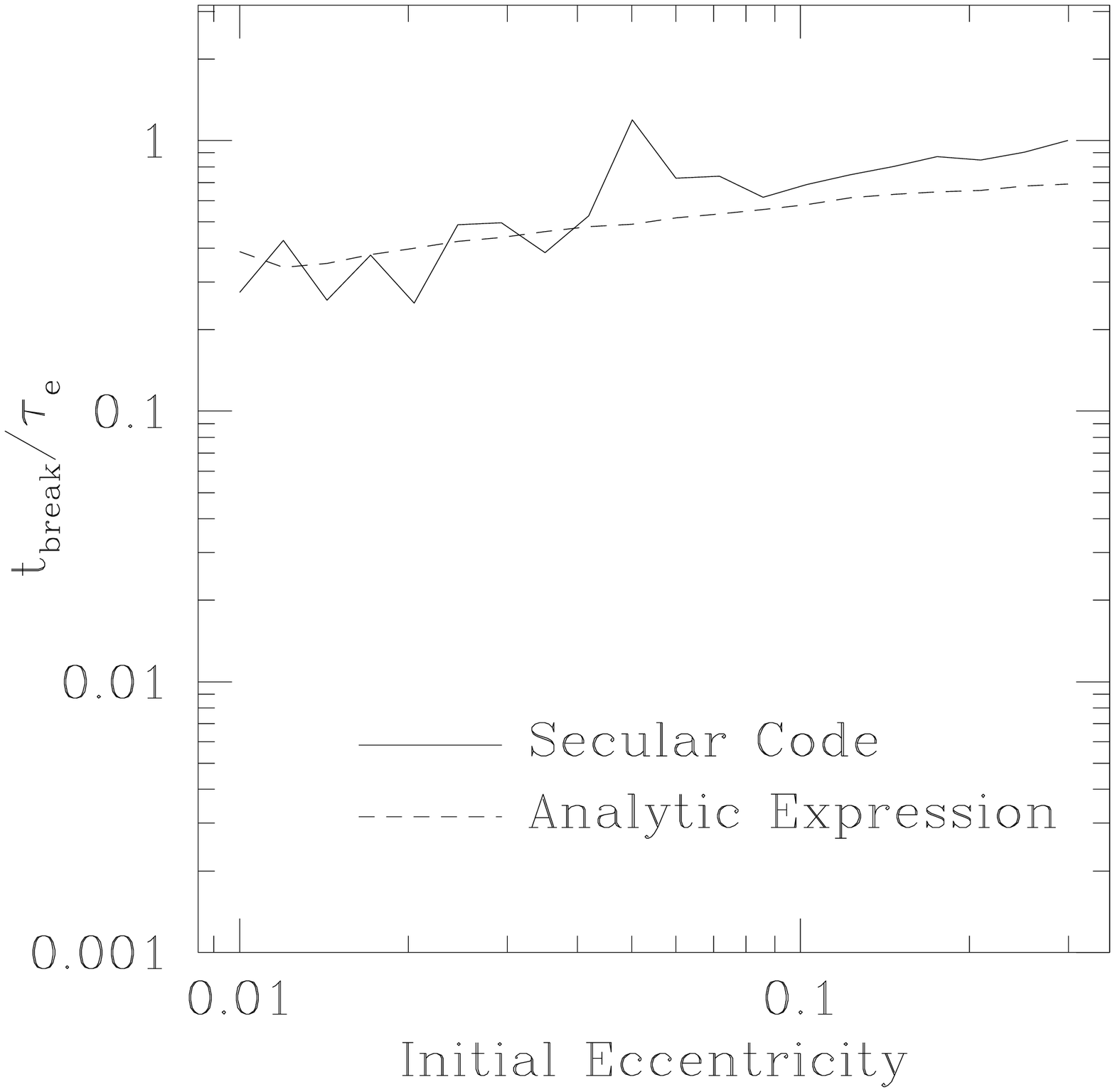}{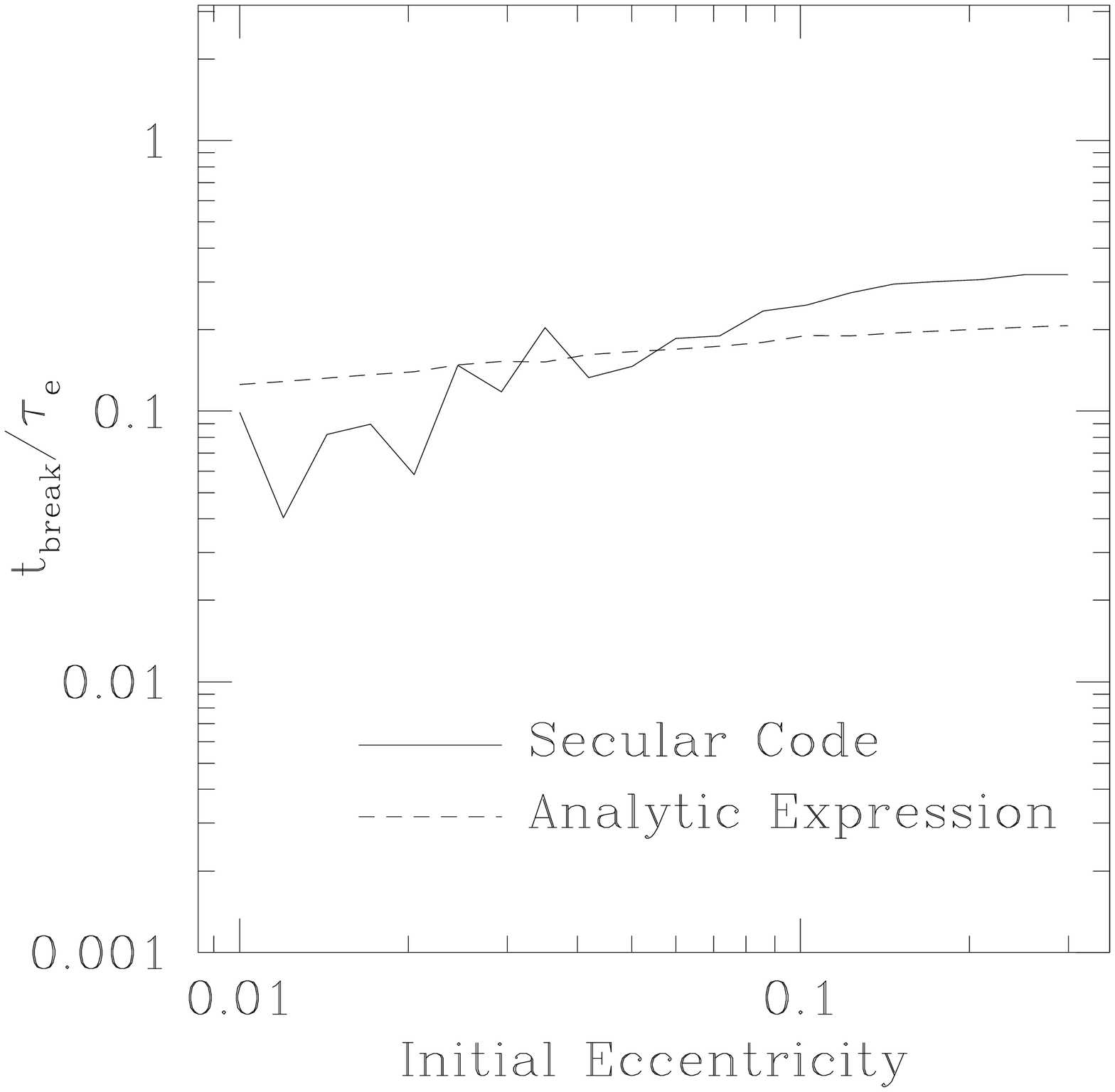}
  \caption{The time required to break the resonance according to
    numerical integrations of the first order secular theory compared
    to the value predicted by equation \ref{tbreakeqn}.  This
    expression takes into account the accelerated circularization
    caused by the change in the dynamics induced by the slight
    dissipation.}
  \label{tbreak}
\end{figure}

\section{Conclusions}

We used the classical disturbing function together with Lagrange's
planetary equations accurate to lowest order in eccentricity to study
resonance breaking through tidal dissipation.  It is possible to
derive analytic expressions for the inner planet's eccentricity when
the resonance is broken and the time required to break the resonance
(equations \ref{ebreakeqn}, \ref{tbreakeqn}).  These expressions are
in good agreement with the secular theory (figure \ref{tbreak})

This is applicable to the 55 Cancri system because the inner planet is
near the regime where tidal effects are starting to play a role and
current fits to the radial velocity curve give orbital elements where
the planets are {\em not} in the resonance.  We argue that the two
planets became trapped in resonance while farther from the star and
then broke the resonance when tidal dissipation began to play a role
in the inner planet's evolution.

\end{document}